\documentclass[runningheads]{llncs}
\usepackage[T1]{fontenc}
\usepackage{graphicx}
\usepackage{bbding}
\usepackage{soul}
\usepackage{bibunits}
\usepackage{bm}
\usepackage{array}
\usepackage{xcolor}
\usepackage{pdflscape}
\usepackage{makecell}
\usepackage{framed,multirow}
\usepackage{threeparttable}
\usepackage[colorlinks,
            anchorcolor=red,
            citecolor=citecolor, 
            linkcolor=linkcolor,
            ]{hyperref}
\usepackage{amssymb}
\usepackage{pifont}
\usepackage{algorithm}
\usepackage{layouts}
\usepackage{listings}
\usepackage{pythonhighlight}
\usepackage{amsmath}
\usepackage{siunitx}
\usepackage{cite}
\makeatletter
\newcommand\footnoteref[1]{\protected@xdef\@thefnmark{\ref{#1}}\@footnotemark}
\makeatother

\definecolor{citecolor}{HTML}{0071BC}
\definecolor{linkcolor}{HTML}{ED1C24}

\DeclareMathOperator*{\argmax}{argmax}

\newcolumntype{P}[1]{>{\centering\arraybackslash}p{#1}}
\newlength\savewidth\newcommand\shline{\noalign{\global\savewidth\arrayrulewidth
  \global\arrayrulewidth 0.8pt}\hline\noalign{\global\arrayrulewidth\savewidth}}
\newcommand{\eg}{\mbox{e.g.,\ }}
\newcommand{\etal}{\mbox{et al.}}

\def\arrvline{\hfil\kern\arraycolsep\vline\kern-\arraycolsep\hfilneg}

\definecolor{Highlight}{HTML}{39b54a}  

\begin{document}

\title{Continual Learning for Abdominal Multi-Organ and Tumor Segmentation}
\titlerunning{Continual Learning for Abdominal Multi-Organ and Tumor Segmentation}

\author{Yixiao Zhang \and
Xinyi Li \and
Huimiao Chen \and \\
Alan L. Yuille \and
Yaoyao Liu\inst{\star} \and
Zongwei Zhou\thanks{Corresponding authors: Yaoyao Liu and Zongwei Zhou (\{yliu538, zzhou82\}@jh.edu)}
}
\authorrunning{Y. Zhang et al.}
%
\institute{Johns Hopkins University \\
{\scriptsize Code:~\url{https://github.com/MrGiovanni/ContinualLearning}
}
}

\maketitle              
\begin{abstract}


The ability to dynamically extend a model to new data and classes is critical for multiple organ and tumor segmentation. However, due to privacy regulations, accessing previous data and annotations can be problematic in the medical domain. This poses a significant barrier to preserving the high segmentation accuracy of the old classes when learning from new classes because of the catastrophic forgetting problem. In this paper, we first empirically demonstrate that simply using high-quality pseudo labels can fairly mitigate this problem in the setting of organ segmentation. Furthermore, we put forward an innovative architecture designed specifically for continuous organ and tumor segmentation, which incurs minimal computational overhead. Our proposed design involves replacing the conventional output layer with a suite of lightweight, class-specific heads, thereby offering the flexibility to accommodate newly emerging classes. These heads enable independent predictions for newly introduced and previously learned classes, effectively minimizing the impact of new classes on old ones during the course of continual learning. We further propose incorporating Contrastive Language–Image Pretraining (CLIP) embeddings into the organ-specific heads. These embeddings encapsulate the semantic information of each class, informed by extensive image-text co-training. The proposed method is evaluated on both in-house and public abdominal CT datasets under organ and tumor segmentation tasks. Empirical results suggest that the proposed design improves the segmentation performance of a baseline model on newly-introduced and previously-learned classes along the learning trajectory.

\end{abstract}

\section{Introduction}
\label{sec:introduction}

Humans inherently learn in an incremental manner, acquiring new concepts over time without forgetting previous ones. In contrast, deep learning models suffer from catastrophic forgetting~\cite{lewandowsky1995catastrophic}, where learning from new data can override previously acquired knowledge. In this context, the class-incremental continual learning problem was formalized by Rebuffi~\etal~\cite{rebuffi2017icarl}, where new classes are observed in different stages, restricting the model from accessing previous data.

The medical domain faces a similar problem: the ability to dynamically extend a model to new classes is critical for multiple organ and tumor segmentation, wherein the key obstacle lies in mitigating `forgetting.' A typical strategy involves retaining some previous data. For instance, Liu~\etal~\cite{liu2022learning} introduced a memory module to store the prototypical representation of different organ categories. However, such methods, reliant on an account of data and annotations, may face practical constraints as privacy regulations could make accessing prior data and annotations difficult~\cite{langlotz2019roadmap}. An alternative strategy is to use pseudo labels generated by previously trained models on new data. Ozdemir~\etal~\cite{ozdemir2018learn,ozdemir2019extending} extended the distillation loss to medical image segmentation. A concurrent study of ours~\cite{ji2023continual} mainly focused on architectural extension, addressing the forgetting problem by freezing the encoder and decoder and adding additional decoders when learning new classes. While these strategies have been alleviating the forgetting problem, they led to tremendous memory costs for model parameters.

Therefore, we identify two main open questions that must be addressed when designing a multi-organ and tumor segmentation framework.
\textbf{Q1:}~Can we relieve the forgetting problem without needing previous data and annotations?
\textbf{Q2:}~Can we design a new model architecture that allows us to share more parameters among different continual learning steps? 

To tackle the above questions, in this paper, we propose a novel continual multi-organ and tumor segmentation method that overcomes the forgetting problem with little memory and computation overhead. \textbf{First}, inspired by knowledge distillation methods in continual learning~\cite{li2017learning,michieli2019incremental,liu2023online,liu2021rmm}, we propose to generate soft pseudo annotations for the old classes on newly-arrived data. This enables us to recall old knowledge without saving the old data. We observe that with this simple strategy, we are able to maintain a reasonable performance for the old classes. \textbf{Second}, we propose image-aware segmentation heads for each class on top of the shared encoder and decoder. These heads allow the use of a single backbone and easy extension to new classes while bringing little computational cost. 
Inspired by Liu~\etal~\cite{liu2023clip}, we adopt the text embedding generated by Contrastive Language–Image Pre-training (CLIP)~\cite{radford2021learning}. CLIP is a large-scale image-text co-training model that is able to encode high-level visual semantics into text embeddings. This information will be an advantage for training new classes with the class names known in advance.

We focus on organ/tumor segmentation because it is one of the most critical tasks in medical imaging~\cite{isensee2021nnu,zhou2022interpreting,zhou2021towards,qu2023annotating}, and continual learning in semantic segmentation is under-explored in the medical domain.
We evaluate our continual learning method using three datasets: BTCV~\cite{landman2015miccai}, LiTS~\cite{bilic2019liver} and JHH~\cite{xia2022felix} (a private dataset at Johns Hopkins Hospital)\footnote{The JHH dataset has 200 abdominal CT scans with per-voxel annotations for 13 organs, three gastrointestinal tracts, and four cardiovascular system structures.}. On the public datasets, the learning trajectory is to first segment 13 organs in the BTCV dataset, then learn to segment liver tumors in the LiTS dataset. On the private dataset, the learning trajectory is to first segment 13 organs, followed by continual segmentation of three gastrointestinal tracts and four cardiovascular system structures.
In our study, we review and compare three popular continual learning baselines that apply knowledge distillation to predictions~\cite{li2017learning}, features~\cite{michieli2019incremental}, and multi-scale pooled features~\cite{douillard2021plop}, respectively. The extensive results demonstrate that the proposed method outperforms existing methods, achieving superior performance in both keeping the knowledge of old classes and learning the new ones while maintaining high memory efficiency. 

\begin{figure}[t]
\centerline{\includegraphics[width=1\columnwidth]{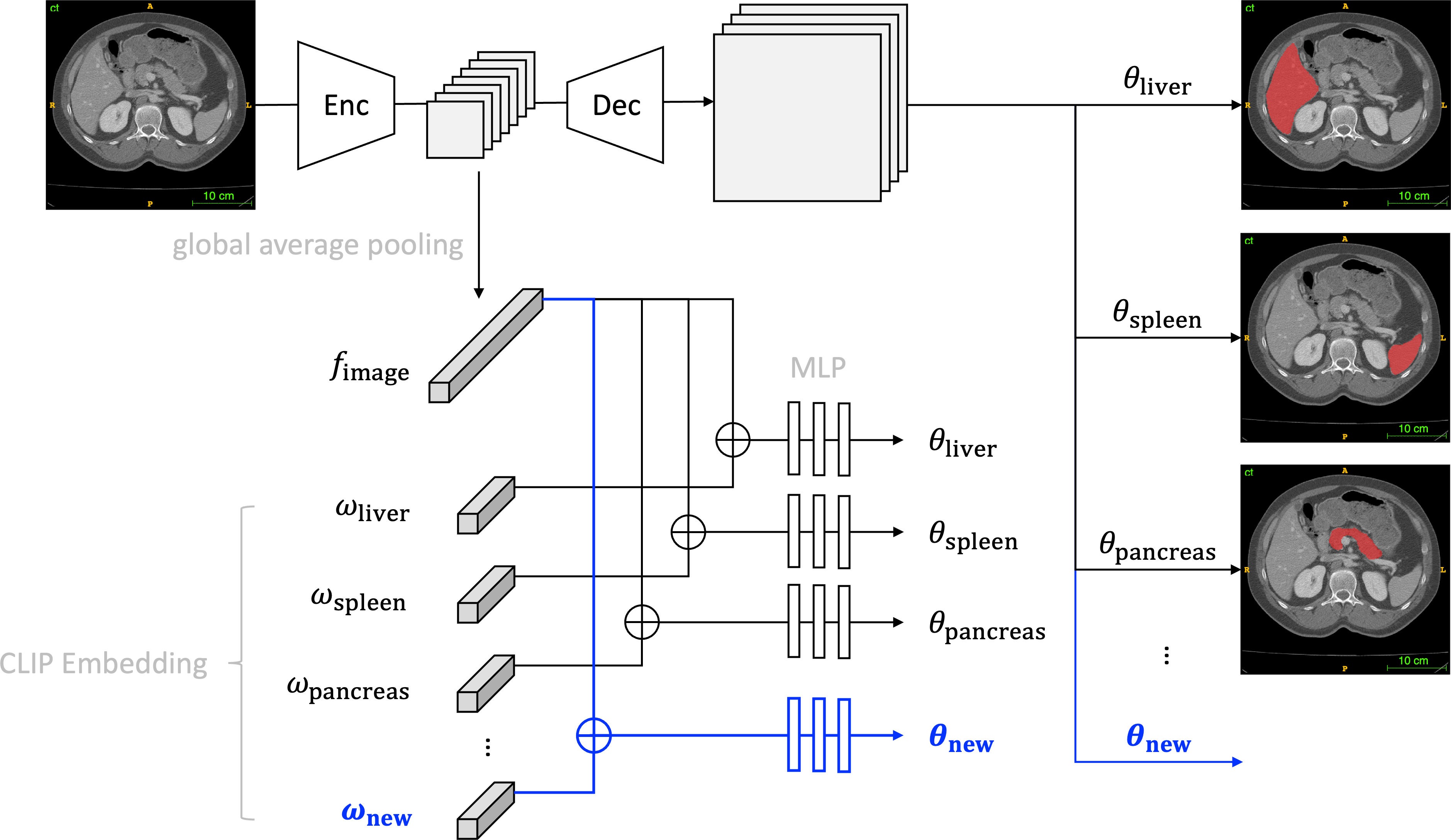}}
    \caption{
    An overview of the proposed method. An encoder (Enc) processes the input image to extract its features, which are then reduced to a feature vector ($f_{\text{image}}$) by a global average pooling layer. This feature vector is subsequently concatenated with a CLIP embedding ($\omega_{\text{class}}$), calculated using the pre-trained CLIP model. Through a series of Multi-Layer Perceptron (MLP) layers, we derive class-specific parameters of convolution kernels ($\theta_{\text{class}}$). These kernels, when applied to the decoder (Dec) feature, yield the mask for the respective class.
    }
\label{fig:method}
\end{figure}

\section{Methodology}
\label{sec:method}
{\sloppy We formulate the continual organ segmentation as follows: given a sequence of partially annotated datasets $\{D_1, D_2, \dots, D_n\}$ each with organ classes $\{C_1, C_2, \dots, C_n\}$, we learn a single multi-organ segmentation model sequentially using one dataset at a time. When training on the $i$-th dataset $D_t$, the previous datasets $\{D_1, \dots, D_{t-1}\}$ are not available. The model is required to predict the accumulated organ labels for all seen datasets $\{D_1, \dots, D_t\}$ :
\begin{align}
\hat{Y}_j &= \argmax_{c \in \mathcal{C}_t} P(Y_j=c|X) \\
\mathcal{C}_t &= \cup_{\tau \leq t} C_\tau
\end{align}
where $j$ is a voxel index, $X$ is an image from $D_t$, $P$ is the probability function that the model learns and $\hat{Y}$ is the output segmentation mask.\par}

\subsection{Pseudo labels for multi-organ segmentation}
In the context of continual organ segmentation, the model's inability to access the previous dataset presents a challenge as it often results in the model forgetting the previously learned classes. In a preliminary experiment, we observed that a segmentation model pre-trained on some organ classes will totally forget the old classes when fine-tuned on new ones. We found the use of pseudo-labeling can largely mitigate this issue and preserve the existing knowledge. Specifically, we leverage the output prediction from the previous learning step $t-1$, denoted as $\hat{Y}_{t-1}$, which includes the old classes $\mathcal{C}_{t-1}$, as the pseudo label for the current step's old classes. For new classes, we still use the ground truth label. Formally, the label $\tilde{L_{t}^{c}}$ for class $c$ in current learning step $t$ can be expressed as:
\begin{align}
\tilde{L_{t}^{c}} &= 
\begin{cases}
L_{t}^{c}   & \text{ if } c \in \mathcal{C}_t-\mathcal{C}_{t-1} \\
\hat{Y}_{t-1}^{c}   & \text{ if } c \in \mathcal{C}_{t-1}
\end{cases}
\end{align}
where $L_{t}^{c}$ represents the ground truth label for class $c$ in step $t$ obtained from dataset $D_{t}$.
By utilizing this approach, we aim to maintain the original knowledge and prevent the model from forgetting the previously learned information while learning the new classes. The following proposed model is trained only with pseudo labeling of old classes without any other distillation or regularization.

\subsection{The proposed multi-organ segmentation model}
In the following, we introduce the proposed multi-organ segmentation model for continual learning. Figure~\ref{fig:method} illustrates the overall framework of the proposed model architecture. It has an encoder-decoder backbone, a set of image-aware organ-specific output heads, and text-driven head parameter generation.

\smallskip\noindent\textbf{\textit{Backbone model:}} 
For continual learning, ideally, the model should be able to learn a sufficiently general representation that would easily adapt to new classes. We use Swin~UNETR~\cite{hatamizadeh2022swin} as our backbone since it exhibits strong performance in self-supervised pre-training and the ability to transfer to various medical image segmentation tasks. Swin~UNETR has Swin~Transformer~\cite{liu2021swin} as the encoder and several deconvolution layers as the decoder. 

\smallskip\noindent\textbf{\textit{Image-aware organ-specific heads:}} 
The vanilla Swin~UNETR has a Softmax layer as the output layer that predicts the probabilities of each class. We propose to replace the output layer with multiple image-aware organ-specific heads. We first use a global average pooling (GAP) layer on the last encoder features to obtain a global feature $f$ of the current image $X$. Then for each organ class $k$, a multilayer perceptron (MLP) module is learned to map the global image feature to a set of parameters $\theta_k$:
\begin{equation}
\theta_k = \text{MLP}_k(\text{GAP}(E(X))),
\end{equation}
where $E(X)$ denotes the encoder feature of image $X$. An output head for organ class $k$ is a sequence of convolution layers that use parameters $\theta_k$ as convolution kernel parameters. These convolution layers are applied to the decoder features, which output the segmentation prediction for organ class $k$:
\begin{equation}
P(Y^k_j = 1 | X, \theta_k) = \sigma(\text{Conv}(D(E(X));\theta_k)),
\end{equation}
where $E$ is the encoder, $D$ is the decoder, $\sigma$ is the Sigmoid non-linear layer and $P(Y^k_j = 1)$ denotes the predicted probability that pixel $j$ belongs to the organ class $k$. The predictions for each class are optimized by Binary Cross Entropy loss. The separate heads allow independent probability prediction for newly introduced and previously learned classes, therefore minimizing the impact of new classes on old ones during continual learning. 
Moreover, this design allows multi-label prediction for cases where a pixel belongs to more than one class (e.g., a tumor on an organ).

\smallskip\noindent\textbf{\textit{Text driven head parameter generation:}} 
We further equip the segmentation heads with semantic information about each organ class. With the widespread success of large-scale vision-language models, there have been many efforts that apply these models to the medical domain~\cite{chen2022multi,huang2021gloria,zhang2022contrastive}. It is suggested that vision-language models could be used for zero-shot learning in the medical domain and recognize novel classes with well-designed prompts~\cite{qin2022medical}. We propose to use CLIP~\cite{radford2021learning} to generate text embeddings for the target organ names. Specifically, we produce the organ name embedding by the pre-trained CLIP text encoder and a medical prompt (e.g., ``a computerized tomography of a [CLS]'', where [CLS] is an organ class name). Then we use the text embeddings $\omega$ together with the global image feature $f$ to generate parameters for the organ segmentation heads:
\begin{equation}
\theta_k = \text{MLP}_k([\text{GAP}(E(X)), \omega_k]),
\end{equation}
where $\omega_k$ is the text embedding for organ class $k$. CLIP embeddings carry high-level semantic meanings and have the ability to connect correlated concepts. Therefore, it guides the MLP module to generate better convolution parameters for each organ class. More importantly, the fixed-length CLIP embedding allows us to adapt the pre-trained model to open-vocabulary segmentation and extend to novel classes.

\smallskip\noindent\textbf{\textit{Difference from Universal Model~\cite{liu2023clip}:}}
For the purpose of continual learning, we improve the original design of Universal Model in the MLP module. 
Unlike Liu~\etal~\cite{liu2023clip}, who utilized a single MLP to manage multiple classes, we allocate an individual and independent MLP to each class. This design significantly mitigates interference among different classes.

\subsection{Computational complexity analysis}
Another key contribution of our work is the reduction of computational complexity in continual segmentation. We compare our proposed model's FLOPs (floating-point operations per second) with the baseline model, Swin UNETR~\cite{hatamizadeh2022swin}. Our model's FLOPs are just slightly higher than Swin~UNETR's, with 661.6 GFLOPs and 659.4 GFLOPs, respectively. This is because we used lightweight output convolution heads with a small number of channels. Ji~\etal~\cite{ji2023continual} proposed a state-of-the-art architecture for medical continual semantic segmentation, which uses a pre-trained and then frozen encoder coupled with incrementally added decoders in each learning step. However, subsequent continual learning steps using this architecture introduce massive computational complexity. For example, Swin~UNETR's decoder alone has 466.08 GFLOPs, meaning that every new learning step adds an additional 466.08 GFLOPs. In contrast, our model only needs to add a few image-aware organ-specific heads for new classes of the new task, with each head consuming only \num{0.12} GFLOPs. As a result, the computational complexity of our model nearly remains constant in continual learning for segmentation, while that of the architecture of Ji~\etal~\cite{ji2023continual} increases linearly to the number of steps.
Compared with ILT~\cite{michieli2019incremental} and PLOP~\cite{douillard2021plop}, since they require feature distillation by inference with an old model while training a new one, they double the FLOPs and are computationally less efficient than the proposed method.

\section{Experiment \& Result}
\label{sec:experiment_result}

\noindent\textbf{\textit{Datasets:}}
We empirically evaluate the proposed model under two data settings: in one setting, both training and continual learning are conducted on the in-house JHH dataset. It has multiple classes annotated, which can be categorized into three groups: the abdominal organs (in which seven classes are learned in step 1: spleen, right kidney, left kidney, gall bladder, liver, postcava, pancreas), the gastrointestinal tract (in which three classes are learned in step 2: stomach, colon, intestine), and other organs (in which four classes are learned in step 3: aorta, portal vein and splenic vein, celiac truck, superior mesenteric artery). The categorization is in accordance with TotalSegmentator~\cite{wasserthal2022totalsegmentator}.
In the other setting, we first train on the BTCV dataset and then do continual learning on the LiTS dataset. The BTCV dataset contains 47 abdominal CT images delineating 13 organs. The LiTS dataset contains 130 contrast-enhanced abdominal CT scans for liver and liver tumor segmentation. We use 13 classes (spleen, right kidney, left kidney, gall bladder, esophagus, liver, stomach, aorta, inferior vena cava, portal vein and splenic vein, pancreas, right adrenal gland, left adrenal gland) from BTCV in step 1 learning and the live tumor from LiTS in step 2 learning. 

\begin{figure}[t]
\centerline{\includegraphics[width=\columnwidth]{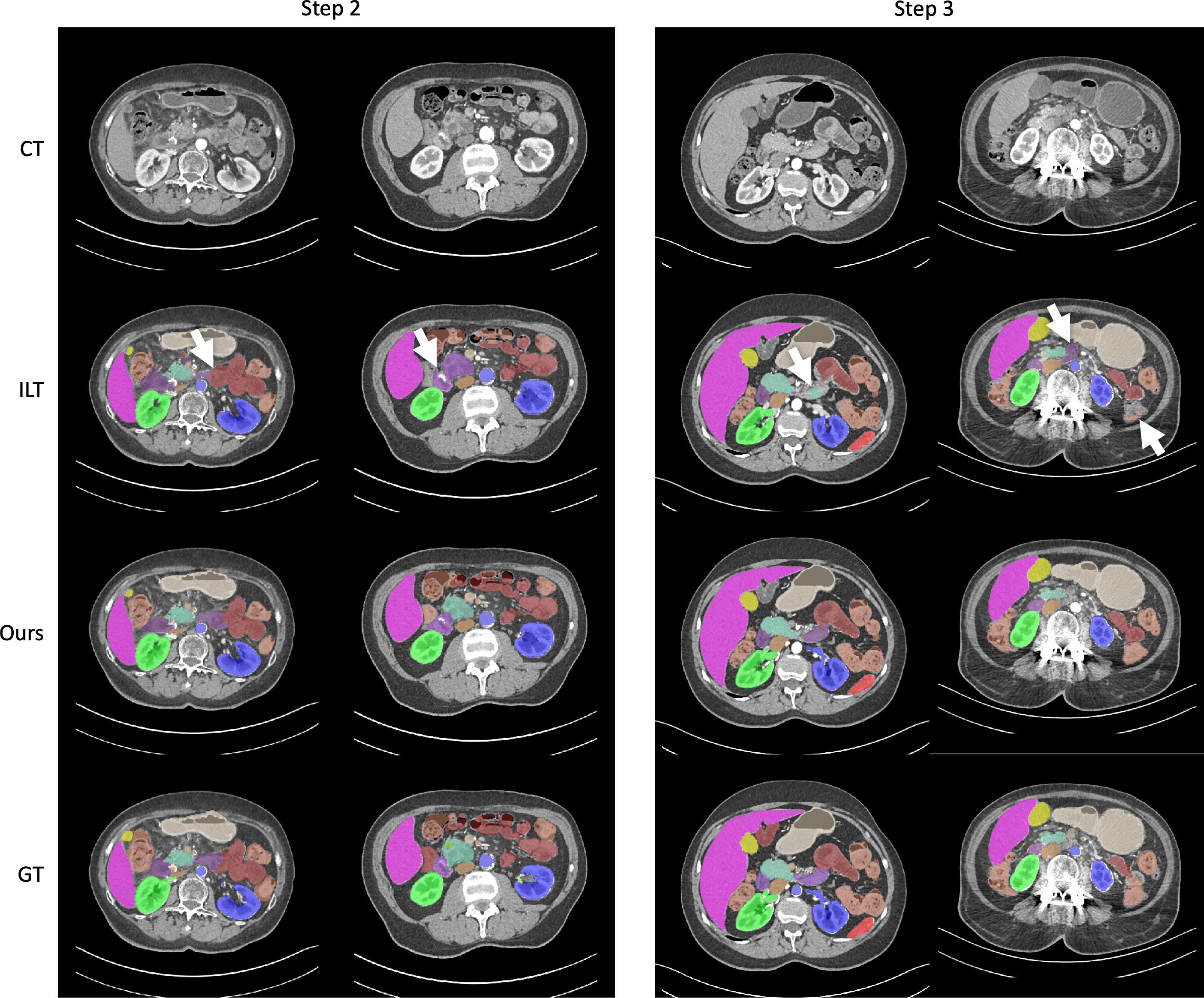}}
    \caption{
    The visualization comparison between our model and the baseline model Swin~UNETR in continual learning steps 2 and 3 on the JHH dataset. 
    }
\label{fig:visual}
\end{figure}

\smallskip\noindent\textbf{\textit{Baselines and metrics:}}
For a fair comparison, all the compared methods use the same Swin~UNETR~\cite{hatamizadeh2022swin} as the backbone, which is the state-of-the-art model in a bunch of medical image segmentation tasks. We compare with three popular continual learning baseline methods that apply knowledge distillation, including LwF~\cite{li2017learning}, ILT~\cite{michieli2019incremental} and PLOP~\cite{douillard2021plop}.
We compare the proposed method with different baseline models using the commonly used Dice score (DSC) metric (the Sørensen–Dice coefficient). In each learning step, we report the average DSC for the classes that are used at the current step as well as the previous steps (\eg in step 2 of the JHH dataset, we report the average dice of the gastrointestinal tracts and the abdominal organs). The dice score at old classes reveals a model's ability to retain its previous knowledge, and the score for the current step classes indicates the model's ability to acquire new knowledge under the regularization of old ones.

\smallskip\noindent\textbf{\textit{Implementation details:}}
The proposed model architecture is trained on new classes with pseudo labeling of old classes. No other distillation techniques are used. We use a lightweight design for the image-aware organ-specific heads. Each head consists of three convolution layers. The number of kernels in the first two layers is 8, and in the last layer is 1. All the compared models are trained using the AdamW optimizer for 100 epochs with a cosine learning rate scheduler. We use a batch size of 2 and a patch size of 96 × 96 × 96 for the training. The initial learning rate is set as $1e^{-4}$, and the weight decay is set as $1e^{-5}$. The version of MONAI\footnote{\url{https://monai.io}} used in our experiments is 1.1.0. Models are trained on NVIDIA TITAN RTX and Quadro RTX 8000 GPUs.

\begin{table}[t]
\scriptsize
\centering
\caption{Benchmark continual learning methods on the JHH dataset.}
    \begin{tabular}{p{0.16\linewidth}|P{0.12\linewidth}P{0.12\linewidth}P{0.12\linewidth}|P{0.12\linewidth}P{0.12\linewidth}|P{0.18\linewidth}}
    \hline
    \multirowcell{2}[0pt][l]{Method} & \multicolumn{3}{c|}{JHH\_organ (7)} & \multicolumn{2}{c|}{JHH\_gastro (3)} & JHH\_cardiac (4) \\
     & Step~1 & Step~2 & Step~3 & Step~2 & Step~3 & Step~3 \\
    \shline
    LwF~\cite{li2017learning} & \textbf{0.891} & 0.777 & 0.767 & 0.530 & 0.486 & 0.360\\
    ILT~\cite{michieli2019incremental} & \textbf{0.891} & 0.775 & 0.776 & 0.653 & 0.480 & 0.484 \\
    PLOP~\cite{douillard2021plop} & \textbf{0.891} & 0.780 & 0.777 & 0.427 & 0.464 & 0.318 \\
    \hline
    Ours & 0.887 & \textbf{0.783} & \textbf{0.787} & \textbf{0.695} & \textbf{0.692} & \textbf{0.636}\\
    \hline
    \end{tabular}
\label{tab:felix_incremental_baseline}
\end{table}

\begin{table}[t]
\scriptsize
\centering
\caption{Benchmark continual learning methods on the public datasets.}
    \begin{tabular}{p{0.16\linewidth}|P{0.12\linewidth}P{0.12\linewidth}|P{0.12\linewidth}}
    \hline
    \multirowcell{2}[0pt][l]{Method} & \multicolumn{2}{c|}{BTCV (13)} & \multicolumn{1}{c}{LiTS (1)} \\
     & Step~1 & Step~2 & Step~2\\
    \shline
    LwF~\cite{li2017learning} & 0.828 & 0.770 & 0.456\\
    ILT~\cite{michieli2019incremental} & 0.828 & 0.786 & 0.335\\
    PLOP~\cite{douillard2021plop} & 0.828 & 0.799 & 0.362\\
    \hline
    Ours & \textbf{0.860} & \textbf{0.817} & \textbf{0.466}\\
    \hline
    \end{tabular}
\label{tab:public_incremental_baseline}
\end{table}

\begin{table}[t]
\scriptsize
\centering
\caption{Ablation study on the JHH dataset.}
    \begin{tabular}{p{0.16\linewidth}|P{0.12\linewidth}P{0.12\linewidth}P{0.12\linewidth}|P{0.12\linewidth}P{0.12\linewidth}|P{0.18\linewidth}}
    \hline
    \multirowcell{2}[0pt][l]{Method} & \multicolumn{3}{c|}{JHH\_organ (7)} & \multicolumn{2}{c|}{JHH\_gastro (3)} & JHH\_cardiac (4) \\
     & Step~1 & Step~2 & Step~3 & Step~2 & Step~3 & Step~3 \\
    \shline
    LwF~\cite{li2017learning} & 0.891 & 0.777 & 0.767 & 0.530 & 0.486 & 0.360\\
    Ours\_1-hot & 0.882 & 0.767 & 0.777 & 0.674 & 0.665 & 0.452 \\
    Ours\_CLIP & \textbf{0.887} & \textbf{0.783} & \textbf{0.787} & \textbf{0.695} & \textbf{0.692} & \textbf{0.636} \\
    \hline
    \end{tabular}
\label{tab:felix_ablation}
\end{table}

\smallskip\noindent\textbf{\textit{Results:}}
The continual segmentation results using the JHH dataset and public datasets are shown in Tables \ref{tab:felix_incremental_baseline} and \ref{tab:public_incremental_baseline}, respectively. Notably, by simply using the pseudo labeling technique (LwF), we are able to achieve reasonably good performance in remembering the old classes (Dice of 0.777 in step 2 and 0.767 in step 3 for abdominal organs in the JHH dataset; Dice of 0.770 in step 2 for BTCV organs). Class-wise DSC scores are in Appendix Tables~\ref{tab:felix_organ}--\ref{tab:btcv_lits}.
All the compared methods use prediction-level or feature-level distillation as regularization. Among them, the proposed method achieves the highest performance in most learning steps. Specifically, the proposed method exhibits the least forgetting in old classes and a far better ability to adapt to new data and new classes. 

To evaluate the proposed model designs, we also conduct the ablation study on the JHH dataset, shown in Table \ref{tab:felix_ablation}. Specifically, we ablate the performance improvement introduced by the organ-specific segmentation heads as well as the CLIP text embeddings. The first line in Table \ref{tab:felix_ablation} shows the performance of the baseline Swin~UNETR model learned with pseudo labeling (LwF). The second row introduces the organ-specific segmentation heads, but uses one-hot embeddings rather than the CLIP text embeddings for each organ. The third row gives the performance of the full method. The results show that by adapting the model to use organ-specific heads as segmentation outputs, we are able to achieve improvement of a large margin (\eg 0.144 in step 2 and 0.179 in step 3 for gastrointestinal tracts). With the application of CLIP text embeddings, we are able to further improves the performance (\eg by a margin of 0.019 in step 2 and 0.027 in step 3 for gastrointestinal tracts). This study validates the effectiveness of the proposed organ-specific segmentation heads and the CLIP text embeddings in the continual organ segmentation task.

Finally, we show the qualitative segmentation results of the proposed method together with the best baseline method ILT on the JHH dataset. We show the results of learning steps 2 and 3 in Figure~\ref{fig:visual}, one case per column and two cases for each step. The visualization demonstrates that the proposed method successfully segments the correct organs while the best baseline method fails throughout the continual learning process.

\section{Conclusion}
\label{sec:conclusion}

In this paper, we propose a method for continual multiple organ and tumor segmentation in 3D abdominal CT images. We first empirically verified the effectiveness of high-quality pseudo labels in retaining previous knowledge. Then, we propose a new model design that uses organ-specific heads for segmentation, which allows easy extension to new classes and brings little computational cost in the meantime. The segmentation heads are further strengthened by utilizing the CLIP text embeddings that encode the semantics of organ or tumor classes.
Numerical results on an in-house dataset and two public datasets demonstrate that the proposed method outperforms the continual learning baseline methods in the challenging multiple organ and tumor segmentation tasks.

\smallskip\noindent\textbf{Acknowledgements.} This work was supported by the Lustgarten Foundation for Pancreatic Cancer Research and partially by the Patrick J. McGovern Foundation Award.
We appreciate the effort of the MONAI Team to provide open-source code for the community.

\newpage
\bibliographystyle{splncs04}
\bibliography{references,zzhou}

\newpage
\appendix

\section*{Appendix}

\begin{table}[h!]
\centering
\scriptsize
\caption{Benchmark continual learning methods on seven classes in the JHH\_organ dataset. We present the Dice score of each class in three continual learning steps. }
\begin{tabular}{p{0.15\linewidth}|P{0.08\linewidth}P{0.1\linewidth}P{0.1\linewidth}P{0.14\linewidth}P{0.07\linewidth}P{0.1\linewidth}P{0.1\linewidth}|P{0.08\linewidth}}
\hline
Method      & Spleen & R Kidney & L Kidney & Gall Bladder & Liver & Postcava & Pancreas & Mean  \\ \shline
\multicolumn{9}{c}{Step 1}   \\ \hline
Swin~UNETR   & 0.942  & 0.947    & 0.932    & 0.827        & 0.960 & 0.799    & 0.829    & \textbf{0.891} \\
Ours\_1-hot & 0.940  & 0.945    & 0.923    & 0.804        & 0.960 & 0.789    & 0.813    & 0.882 \\
Ours\_clip        & 0.945  & 0.943    & 0.931    & 0.806        & 0.960 & 0.781    & 0.843    & 0.887 \\ \hline
\multicolumn{9}{c}{Step 2}   \\ \hline
LwF         & 0.931  & 0.934    & 0.913    & 0.446        & 0.940 & 0.752    & 0.527    & 0.777 \\
ILT         & 0.927  & 0.930    & 0.914    & 0.459        & 0.932 & 0.749    & 0.518    & 0.775 \\
PLOP        & 0.923  & 0.934    & 0.918    & 0.473        & 0.938 & 0.754    & 0.518    & 0.780 \\
Ours\_1-hot & 0.899  & 0.926    & 0.907    & 0.468        & 0.934 & 0.737    & 0.495    & 0.767 \\
Ours\_clip        & 0.933  & 0.930    & 0.918    & 0.453        & 0.939 & 0.756    & 0.555    & \textbf{0.783} \\ \hline
\multicolumn{9}{c}{Step 3}   \\ \hline
LwF         & 0.926  & 0.935    & 0.922    & 0.457        & 0.936 & 0.757    & 0.437    & 0.767 \\
ILT         & 0.939  & 0.942    & 0.924    & 0.452        & 0.944 & 0.757    & 0.476    & 0.776 \\
PLOP        & 0.933  & 0.939    & 0.923    & 0.461        & 0.943 & 0.756    & 0.489    & 0.777 \\
Ours\_1-hot & 0.922  & 0.931    & 0.909    & 0.452        & 0.926 & 0.759    & 0.537    & 0.777 \\
Ours\_clip        & 0.930  & 0.934    & 0.923    & 0.425        & 0.940 & 0.767    & 0.592    & \textbf{0.787} \\ \hline
\end{tabular}
\label{tab:felix_organ}
\end{table}

\begin{table}[h!]
\centering
\scriptsize
\caption{Benchmark continual learning methods on three classes in the JHH\_gastro dataset. We present the Dice score of each class in two continual learning steps.}
\begin{tabular}{p{0.15\linewidth}|P{0.15\linewidth}|P{0.15\linewidth}|P{0.15\linewidth}|P{0.15\linewidth}}
\hline
Method & Stomach & Instine & Colon & Mean \\ \shline
\multicolumn{5}{c}{Step 2} \\
\hline
LwF & 0.859 & 0.104 & 0.628 & 0.530 \\
ILT & 0.854 & 0.488 & 0.617 & 0.653 \\
PLOP & 0.804 & 0 & 0.477 & 0.427 \\
Ours\_1-hot & 0.859 & 0.538 & 0.625 & 0.674 \\
Ours\_clip & 0.859 & 0.561 & 0.664 & \textbf{0.695} \\
\hline
\multicolumn{5}{c}{Step 3} \\
\hline
LwF & 0.867 & 0 & 0.590 & 0.486 \\
ILT & 0.869 & 0 & 0.573 & 0.480 \\
PLOP & 0.852 & 0 & 0.539 & 0.464 \\
Ours\_1-hot & 0.844 & 0.519 & 0.631 & 0.665 \\
Ours\_clip & 0.862 & 0.560 & 0.652 & \textbf{0.692} \\ 
\hline
\end{tabular}
\label{tab:felix_gas}
\end{table}

\begin{table}[h!]
\centering
\scriptsize
\caption{Benchmark continual learning methods on four classes in the JHH\_cardiac dataset. We present the Dice score of each class in the final continual learning step.}
\begin{tabular}{p{0.15\linewidth}|P{0.15\linewidth}|P{0.15\linewidth}|P{0.15\linewidth}|P{0.15\linewidth}|P{0.15\linewidth}}
\hline
Method & Aorta & Veins & Celiac & Sma & Mean \\
\shline
\multicolumn{6}{c}{Step 3} \\
\hline
LwF & 0.840 & 0.560 & 0 & 0.040 & 0.360 \\
ILT & 0.862 & 0.531 & 0.007 & 0.537 & 0.484 \\
PLOP & 0.856 & 0.416 & 0 & 0 & 0.318 \\
Ours\_1-hot & 0.690 & 0.497 & 0.424 & 0.197 & 0.452 \\
Ours\_clip & 0.819 & 0.581 & 0.602 & 0.543 & \textbf{0.636} \\
\hline
\end{tabular}
\label{tab:felix_incremental_baseline_appendix}
\end{table}

\begin{table}[h!]
\centering 
\scriptsize
\caption{Benchmark continual learning methods from the BTCV to LiTS datasets. We present the Dice score of each class in two continual learning steps. }
\begin{tabular}{p{0.16\linewidth}|P{0.1\linewidth}P{0.1\linewidth}P{0.1\linewidth}P{0.16\linewidth}P{0.11\linewidth}P{0.1\linewidth}P{0.1\linewidth}}
\hline
\multicolumn{8}{c}{Step 1} \\
\shline
 & Spleen & R kidney & L kidney & Gall Bladder & Esophagus & Liver & Stomach \\
\hline
Swin~UNETR & 0.955 & 0.819 & 0.927 & 0.844 & 0.718 & 0.969 & 0.896 \\
Ours & 0.952 & 0.917 & 0.922 & 0.840 & 0.720 & 0.966 & 0.886 \\ 
\hline
 & Aorta & IVC & Veins & Pancreas & R gland & L gland & Mean \\ 
\hline
Swin~UNETR & 0.801 & 0.877 & 0.735 & 0.826 & 0.743 & 0.654 & 0.828 \\
Ours & 0.903 & 0.902 & 0.812 & 0.845 & 0.755 & 0.758 & \textbf{0.860} \\
\hline
\multicolumn{8}{c}{Step 2} \\
\shline
 & Spleen & R kidney & L kidney & Gall Bladder & Esophagus & Liver & Stomach \\
\hline
LwF & 0.941 & 0.745 & 0.845 & 0.796 & 0.690 & 0.935 & 0.852 \\
ILT & 0.950 & 0.687 & 0.913 & 0.750 & 0.692 & 0.960 & 0.857 \\
PLOP & 0.942 & 0.778 & 0.908 & 0.823 & 0.690 & 0.959 & 0.883 \\
Ours & 0.941 & 0.860 & 0.872 & 0.728 & 0.690 & 0.955   & 0.862 \\
\hline
 & Aorta & IVC & Veins & Pancreas & R gland & L gland & Mean \\
\hline
LwF & 0.847 & 0.784 & 0.628 & 0.812 & 0.671 & 0.468 & 0.770 \\
ILT & 0.774 & 0.849 & 0.757 & 0.812 & 0.692 & 0.527 & 0.786 \\
PLOP & 0.803 & 0.819 & 0.634 & 0.822 & 0.708 & 0.612 & 0.799 \\
Ours & 0.909 & 0.872 & 0.742 & 0.808 & 0.703 & 0.683 & \textbf{0.817} \\
\hline
\end{tabular}
\label{tab:btcv_lits}
\end{table}

\end{document}